\documentclass{iopart}

\usepackage{epsfig}

\newcommand{\fourfigs}[4]{
  \begin{center}
    \begin{tabular}{cc}
      \epsfig{file=#1, width=0.465\textwidth} &
      \epsfig{file=#2, width=0.465\textwidth} \\
      \mbox{\sl(a)} & \mbox{\sl(b)} \\ 
      \epsfig{file=#3, width=0.465\textwidth} &
      \epsfig{file=#4, width=0.465\textwidth} \\
      \mbox{\sl(c)} & \mbox{\sl(d)}
    \end{tabular}
  \end{center}
  }

\begin{document}

\jl{6}   % Submit to CQG
%\submitted
\newcommand{\eqn}[1]{(\ref{eqn:#1})}
\newcommand{\fig}[2]{\ref{fig:#1}{\sl #2}}
\def\order{{\cal O}}

%\draft
%\begin{flushright}  LAUR-98-****  \end{flushright}
\title[]{Simplicial Brill wave initial data}

\author{Adrian P. Gentle\footnote[1]{Email address adrian@lanl.gov.}}

\address{Department of Mathematics and Statistics, \\
  Monash University, Clayton Victoria 3168, Australia}
\address{and}
\address{Theoretical Division (T-6, MS B288),\\ Los Alamos National 
  Laboratory, Los Alamos, NM 87545, USA}

\begin{abstract}
Regge calculus is used to construct initial data for vacuum
axisymmetric Brill waves at a moment of time symmetry.  We argue that
only a tetrahedral lattice can successfully reproduce the continuum
solution, and develop a simplicial axisymmetric lattice based on the
co-ordinate structure of the continuum metric.  This is used to
construct initial data for Brill waves in an otherwise flat
spacetime, and for the distorted black hole spacetime of Bernstein.
These initial data sets are shown to be second order accurate
approximations to the corresponding continuum solutions.
\end{abstract}
\pacs{}

%\maketitle
\setcounter{footnote}{0}

% --- BODY -----------------------------------------------------------

\section{Introduction}

Since it's inception in 1961, Regge calculus \cite{regge61} has been
extensively studied in highly symmetric spacetimes, for which
corresponding exact solutions of Einstein's equations are often
available.  A complete review and bibliography of this early work is
provided by Williams and Tuckey \cite{williams92}.

Following the realization that a fully decoupled, parallelizable,
three-plus-one dimensional evolution scheme occurs naturally within
Regge calculus \cite{sorkin75,tuckey92,committee}, this simplicial
approach to gravity is now on the verge of tackling physically
interesting and dynamic problems.  The first tentative step along this
path was recently completed, with the successful application of
simplicial Regge calculus to the Kasner spacetime in
$(3+1)$-dimensions \cite{gentle98}.

A vital precursor to the evolution problem in any general relativistic
simulation is the construction of consistent initial data.  Gentle and
Miller \cite{gentle98} present a general prescription for the
calculation of two-surface initial data for the Regge lattice,
although their approach has only been applied to the Kasner cosmology.
As a prelude to the construction of simplicial initial data for
complex spacetimes, in this paper we consider the restricted case of
vacuum, axisymmetric initial data at a moment of time symmetry. This
is the first step towards our goal, and provides a benchmark against
which future, fully four-dimensional, initial data may be compared.

We construct vacuum initial data for the pure Brill wave spacetime
\cite{brill59}, and for Brill waves in a black hole spacetime --- the
``distorted black holes'' first considered by Bernstein
\cite{bernstein93}.  Standard finite-difference techniques have been
used to study pure Brill wave spacetimes by Eppley \cite{eppley77},
Miyama \cite{miyama81}, Holz \etal \cite{holz93} and Alcubierre \etal
\cite{alcubierre98}.  The axisymmetric Brill wave plus black hole
spacetime has been extensively investigated by Bernstein
\cite{bernstein93}, and more recently, a generalization to
$(3+1)$-dimensions has been considered by Camarda \cite{camarda98}.

Dubal \cite{dubal89a} constructed pure Brill wave initial data using
Regge calculus, with a lattice based on prisms rather than simplices.
A similar approach was taken in an earlier version of the current work
\cite{gentle96}.  We will consider the advantages and disadvantages of
using a prism-based lattice, and argue that the prism-based
construction cannot, in general, satisfactorily reproduce the
continuum solution, even in axisymmetry.

We begin in section \ref{sec:continuum} with a short survey of the
important results from the continuum theory. In section
\ref{sec:regge_ivp} we discuss the initial value problem at a moment
of time symmetry in Regge calculus, and then proceed in section
\ref{sec:prism} to examine an axisymmetric lattice built from prisms.
In section \ref{sec:simplicial} we turn to a fully simplicial lattice,
and finally, in section \ref{sec:convergence}, we analyze the
convergence of the simplicial initial data to the continuum solution.

\section{The time symmetric initial value problem}
\label{sec:continuum}

The vacuum constraint equations of General Relativity may be written
in the form (see, for example, Misner, Thorne and Wheeler \cite{mtw})
\begin{eqnarray}
  \label{eqn:constraints}
  R  + \left( \tr K\right) ^2 + K_{ab}K^{ab}    & = & 0 \\
  \nabla_b \left( K^{ab} - \gamma^{ab}\, \tr K \right) & = & 0
\end{eqnarray}
where $R^a_{\ bcd}$ is the intrinsic curvature tensor on the
spacelike hypersurface, and $R$ is the Ricci scalar.  The
extrinsic curvature $K_{ab}$ determines the embedding of this slice in
the spacetime.  At a moment of time symmetry $K_{ab}=0$, and the
constraints reduce to vanishing of the scalar three-curvature:
\begin{equation}  \label{eqn:continuum_ivp}
  R = 0.
\end{equation}

The standard approach to the solution of this equation involves a
conformal decomposition of the three-metric, such that
\begin{equation}
  \label{eqn:conformal}
  \gamma _{ab} = \psi^4 \bar \gamma _{ab}
\end{equation}
which recasts equation \eqn{continuum_ivp} in the form
\begin{equation}
  \label{eqn:hamiltonian_symmetry}
  \nabla ^2 \psi = \frac{1}{8}\, \psi \bar R
\end{equation}
which is linear in the conformal factor $\psi$.  The base metric $\bar
\gamma_{ab}$ is used to calculate the base scalar curvature $\bar R$.
The base three-geometry may be freely specified, although the choice
may restrict the class of possible solutions.  Solving this single
linear elliptic partial differential equation yields the full initial
data set at a moment of time symmetry.

\subsection{Pure gravitational radiation}
\label{sec:pure_brill}

In this section we introduce the axisymmetric vacuum initial value
problem posed by Brill \cite{brill59}.  Conformal decomposition using
a flat base metric leads only to trivial solutions.  The approach
taken by Brill \cite{brill59} was to introduce a metric of the
form
\begin{equation}
  \label{eqn:brill}
  ds^2 = \psi^4 \left\{ e^{2q} ( d\rho ^2 +dz^2 ) +
    \rho^2 \, d\phi^2 \right\},
\end{equation}
where the arbitrary function $q(\rho,z)$ can be considered the
distribution of gravitational wave amplitude \cite{wheeler64}.  Brill
showed that $q(\rho,z)$ must satisfy the boundary conditions
\begin{displaymath}       
  q(0,z) = 0, \quad q_\rho(0,z) = 0, \quad \hbox{and} \quad q_z (\rho,0)
  = 0,
\end{displaymath}
together with the condition that $q = {\cal O}(r^{-2})$ as $r\rightarrow
\infty$, to ensure that the hypersurface has an asymptotically well
defined mass.  With this choice of background metric, the Hamiltonian
constraint \eqn{hamiltonian_symmetry} takes the form
\begin{equation}  \label{eqn:brill_hamiltonian}
  \bigtriangledown ^2 \psi = - \frac {\psi}{4} \left( \frac {\partial
      ^2 q}{\partial \rho ^2} + \frac {\partial ^2 q}{\partial z ^2}
  \right)
\end{equation}
which is solved for $\psi (\rho,z)$ once $q(\rho,z)$ is given.

\begin{figure}
  \begin{center}
    \epsfig{file=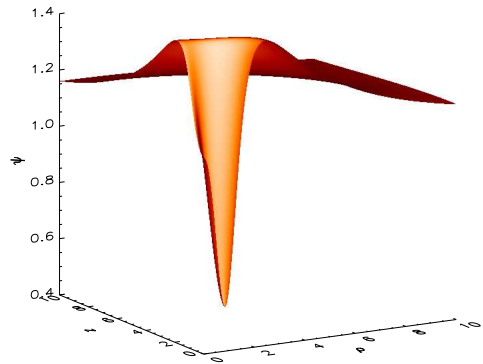, width=0.6\textwidth} 
  \end{center}
  \caption{The conformal factor $\psi$ for Brill wave initial data,
    calculated on a $601\times601$ grid, using a centred finite
    difference approximation to equation \eqn{brill_hamiltonian}.  The
    Eppley form of $q(\rho,z)$ (equation \eqn{brill_q}) was used, with
    wave amplitude $a=10$, and the outer boundaries were placed at $\rho
    = 20$ and $z=20 $.}
  \label{fig:continuum_brill}
\end{figure}

To allow comparison with Eppley \cite{eppley77}, we choose $q(\rho,z)$
to be of the form
\begin{equation}   \label{eqn:brill_q}
q = \frac {a \rho ^2 }{ 1+ r^n} \qquad \hbox{where} \quad r^2 = \rho
^2 + z^2,
\end{equation}
and the boundary conditions on $q$ imply that $n \ge 4$.  In the
remainder of this paper we set $n=5$.  The single remaining parameter,
the wave amplitude $a$, is arbitrary.

To obtain an axisymmetric, asymptotically flat solution which is 
reflection symmetric on the $z=0$ axis, we use boundary conditions on
$\psi$ of the form
\begin{equation}  \label{eqn:brill_boundary1}
  \frac {\partial \psi}{\partial \rho}(0,z) = 0, \qquad \quad
  \frac {\partial \psi}{\partial z}(\rho,0) = 0,  
\end{equation}
together with a Robin outer boundary condition,
\begin{equation}
  \label{eqn:brill_boundary2}
  \frac {\partial \psi}{\partial r} = \frac{ 1-\psi
    } {r} \quad \hbox{as} \quad r \rightarrow \infty.
\end{equation}

Figure \fig{continuum_brill}{a} shows the solution of equation
\eqn{brill_hamiltonian} obtained using a centred finite difference
approximation to both the equation and boundary conditions.  The
solution was calculated on a $601\times 601$ grid with a Brill wave
amplitude $a=10$.

% -----------------------------------------------------------------

\subsection{Distorted black hole initial data}
\label{sec:distorted_bh}

We now turn to the Brill wave plus black hole spacetime, investigated
in great detail by Bernstein \cite{bernstein93}.  
As before, a perturbation is introduced onto the background metric,
and the conformal factor is calculated from the single initial value
equation \eqn{hamiltonian_symmetry}.  Note that we solve the full
initial value problem for the wave and black hole together.  This is
not a perturbation solution.
 
Mirroring the pure Brill wave spacetime of section
\ref{sec:pure_brill}, we write the physical metric on the initial
surface in the form
\begin{equation}
  \label{eqn:black_hole_brill}
  ds^2 = \psi ^4 \left\{ \mbox{e}^{2q} \left( d\eta ^2 +d\theta ^2\right)  +
    \sin ^2 \theta \, d\phi^2 \right\},
\end{equation}
where the exponential radius co-ordinate $\eta$ is defined from $\rho
= m \exp (\eta) /2$.  The ``mass'' $m$ is that of the black hole
alone, obtained by setting $q(\eta,\theta)=0$.  In $(\eta,\theta)$
co-ordinates, the isolated black hole solution takes the form
\begin{equation}
  \label{eqn:conformal_black_hole}
  \psi _{bh} = \sqrt{2m} \, \cosh \left( \eta /2 \right).  
\end{equation}

The Hamiltonian constraint \eqn{hamiltonian_symmetry} evaluated on the
black hole plus Brill wave metric, equation \eqn{black_hole_brill},
yields the linear constraint
\begin{equation}  \label{eqn:brill_bh_hamiltonian}
  \bigtriangledown ^2 \psi = - \frac {\psi}{4} \left( \frac {\partial
      ^2 q}{\partial \eta ^2} + \frac {\partial ^2 q}{\partial \theta ^2}
  - 1 \right)
\end{equation}
which is again solved for the conformal factor $\psi$ once the
perturbation $q(\eta,\theta)$ is given.  
We choose boundary conditions of the form
\begin{equation}
  \label{eqn:hole_boundary1}
  \frac {\partial \psi}{\partial \eta}(0,\theta) = 0, \mbox{\qquad}
  \frac {\partial \psi}{\partial \theta}(\eta,0) = 0, \mbox{\qquad}
  \frac {\partial \psi}{\partial \theta}(\eta,\frac{\pi}{2} ) = 0,
\end{equation}
on the ``inner'' boundaries, where $\eta=0$ corresponds to an
Einstein-Rosen bridge at $r=m/2$.  These represent reflection symmetry
on the $\theta=0$, $\theta=\pi/2$ and $\eta=0$ boundaries. 
The outer boundary condition is again applied using the Robin
approach, adapted to the exponential radial co-ordinate $\eta$. 
Examining the asymptotic form of the black hole solution
yields the condition
\begin{equation}
  \label{eqn:hole_boundary2}
  \frac {\partial \psi}{\partial \eta} + \frac{\psi}{2}
  = \sqrt{\frac{m}{2}} \, {e}^{\eta/2},
\end{equation}
on the outer boundary.

\begin{figure}
  \fourfigs  
  {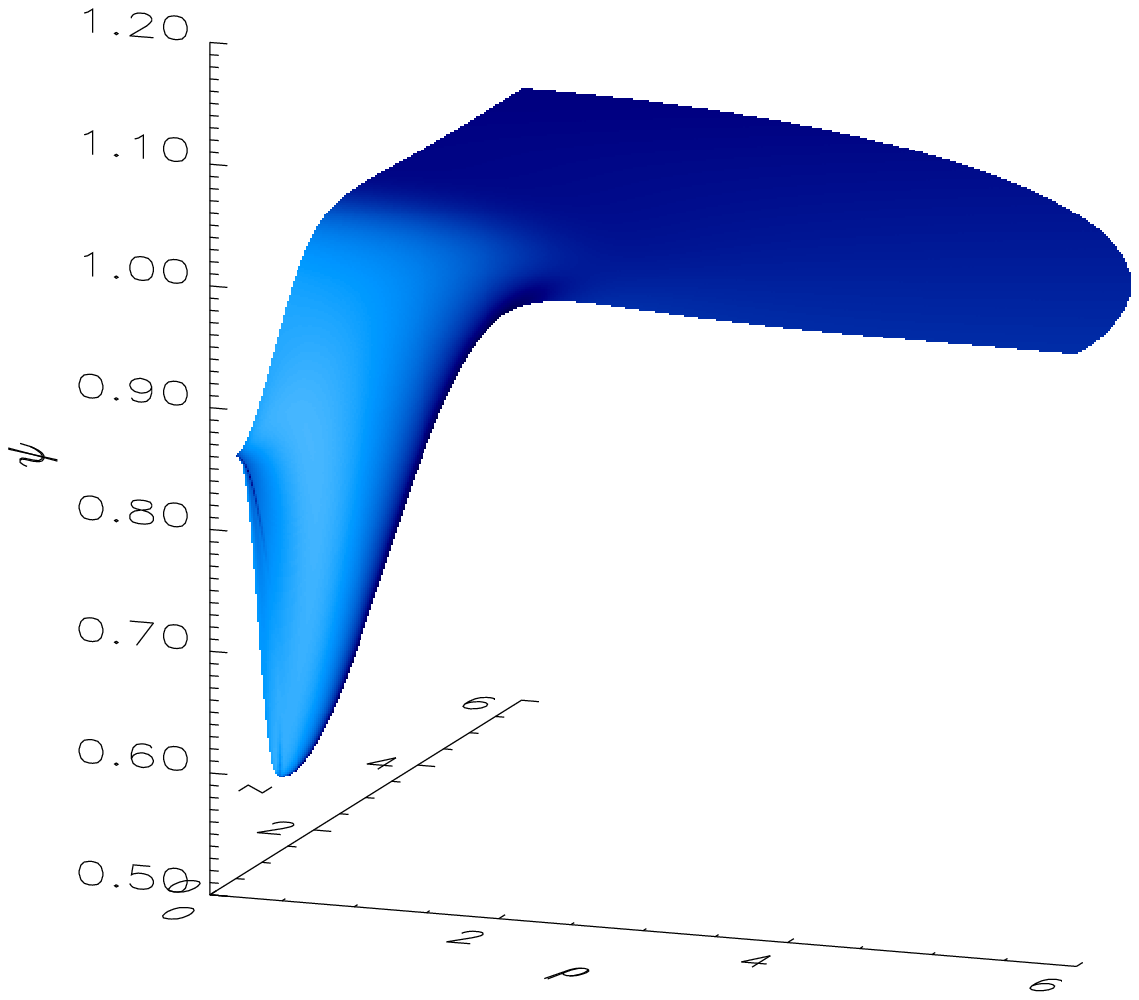}  {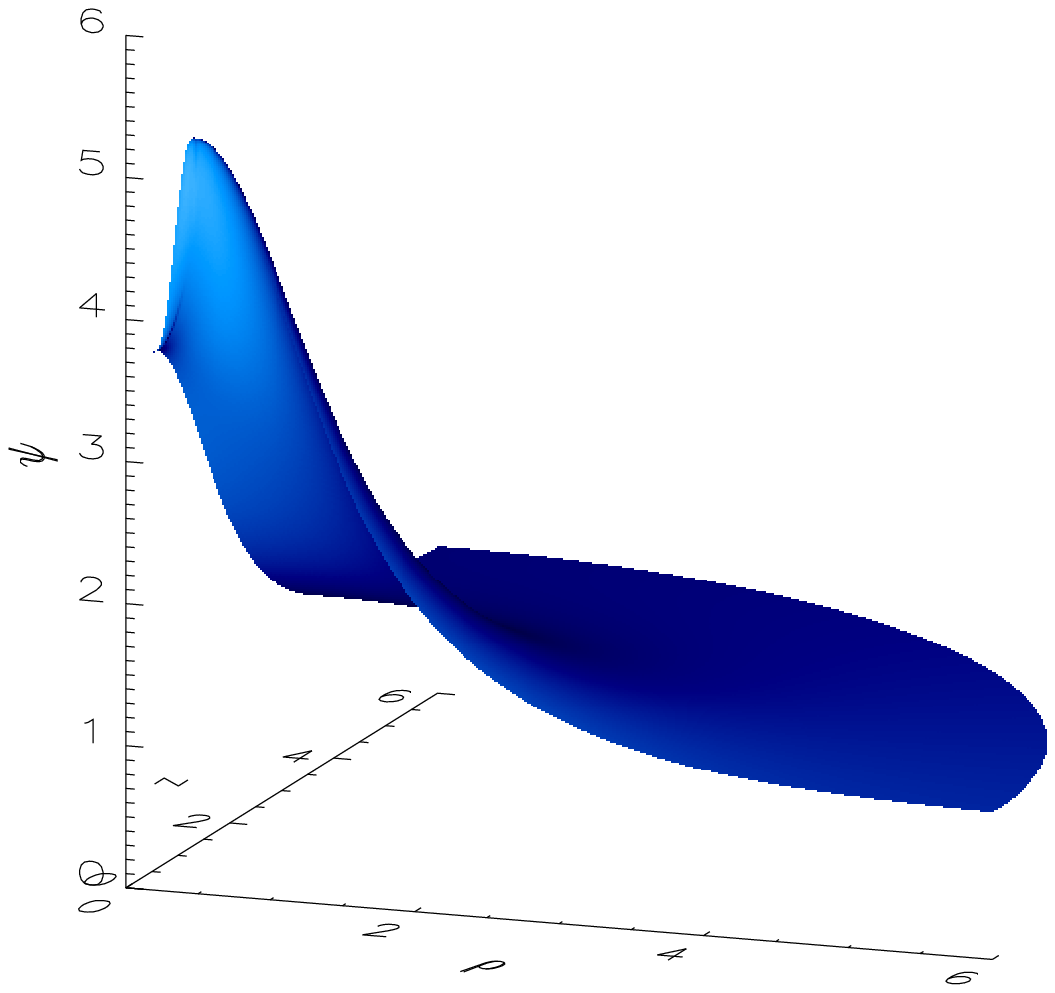}  {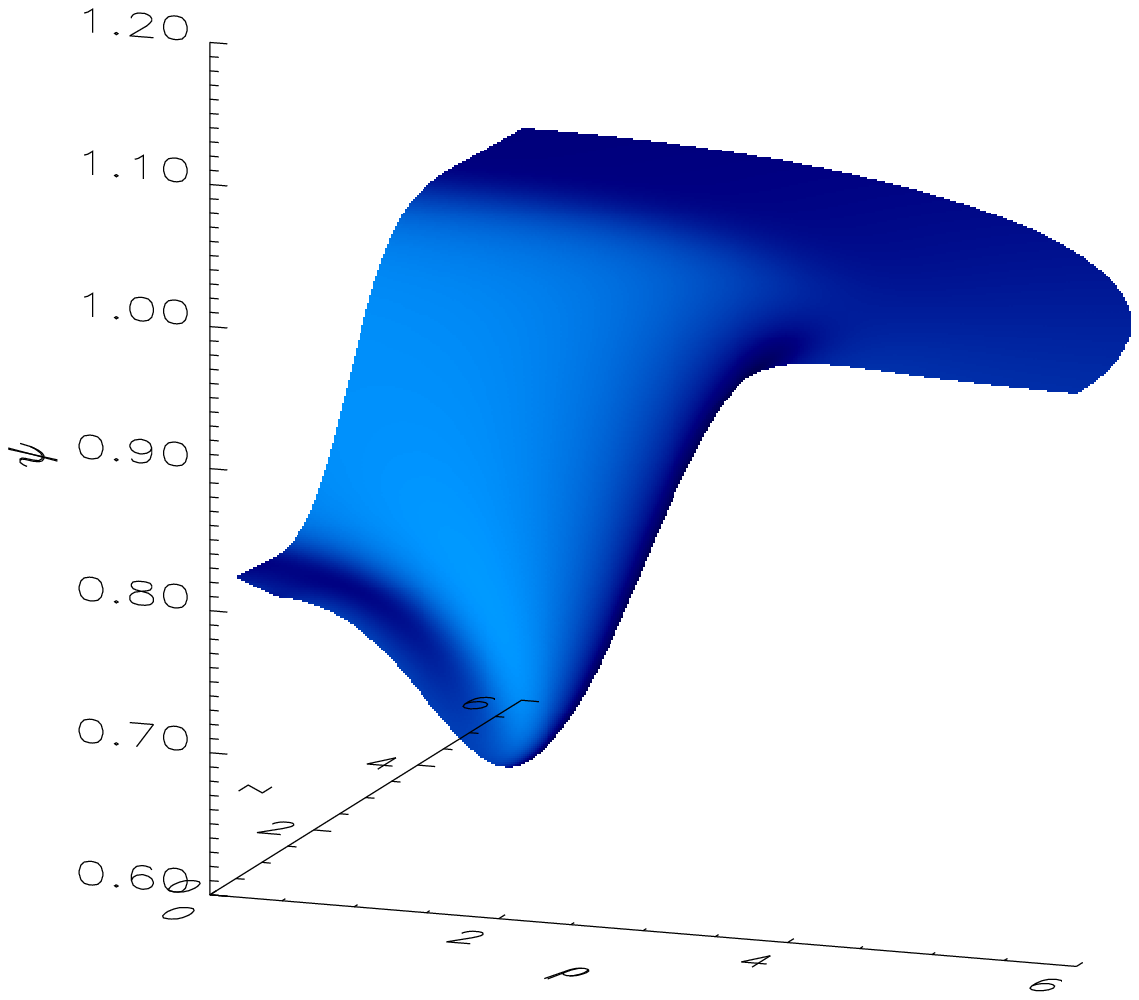}  {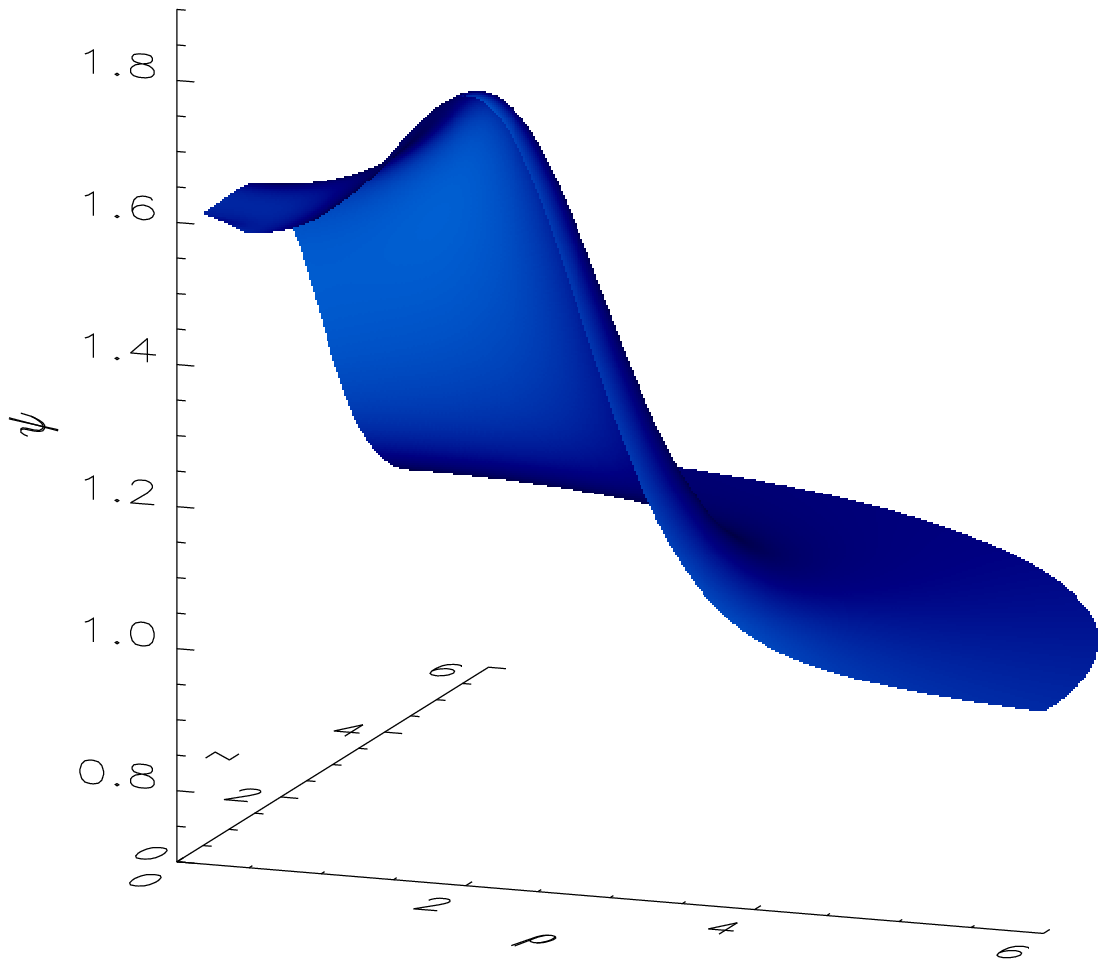}
  \caption{The continuum solution for the black hole plus Brill wave
    spacetime.  The conformal factor is plotted as a ratio of the
    black hole solution ($\psi/\psi_{bh}$), with the results
    calculated on a $601\times601$ grid with $m=1$.  An artificial
    Cartesian projection is used, and this is defined in the text.
    The choice of parameters are (a) $(1,0,1)$, (b) $(-1,0,1)$, (c)
    $(1,2,1)$ and (d) $(-1,2,1)$.}
    \label{fig:continuum_plots}
\end{figure}

As a test of the axisymmetric code, we solve equation
\eqn{brill_bh_hamiltonian} using a second-order finite difference
approximation with $q(\eta,\theta)=0$, together with the boundary
conditions above.  We expect that the solution $\psi_{ij}$ will
converge as the second power of the grid spacing to the analytic black
hole solution, given by equation \eqn{conformal_black_hole}.  We have
performed such a calculation, and find that the axisymmetric numerical
solution does indeed converge as the second power of the grid spacing
to the black hole solution.  This provides some confidence in the the
continuum code, which is vital since it will be used later to
benchmark the Regge calculations.

Returning to the construction of distorted black hole initial data, we
follow Bernstein and choose $q(\eta, \theta )$ to be of the form
\begin{equation}
  \label{eqn:hole_q}
q = a \sin ^2 \theta \left\{ e^{-g_+^2} + e^{-g_-^2} \right\}  
\end{equation}
with $g_{\pm} = \left( \eta \pm b \right) / \omega $, which contains
the set of free parameters $(a,b,\omega)$.

The solution of the Hamiltonian constraint \eqn{brill_bh_hamiltonian}
is shown in figure \fig{continuum_plots}{} for several illustrative
choices of the free parameters.  The figure shows the ratio of the
conformal factor to the black hole solution, $\psi /\psi_{bh}$.  The
outer boundary was placed at $\eta=6$, and the results are plotted on
a pseudo-Cartesian grid defined by
\begin{eqnarray}
  \label{eqn:cartesian_coords}
  x & = & (\eta+\eta_0)\sin \theta  \nonumber \\
  y & = & (\eta+\eta_0)\cos \theta  \nonumber 
\end{eqnarray}
where $\eta_0 = \frac{1}{2}$ is used to display an artificial throat
in $(\eta, \theta)$ coordinates.  These results compare well with the
calculations of Bernstein \cite{bernstein93}, who also used a second
order finite-difference approximation to the Hamiltonian constraint.

% -----------------------------------------------------------------

\section{The Regge initial value problem at a moment of time symmetry}
\label{sec:regge_ivp}

We found in section \ref{sec:continuum} that the continuum initial
value problem at a moment of time symmetry reduces to the vanishing of
the scalar three-curvature intrinsic to the hypersurface.  By using
this knowledge, the full two-slice Regge initial value problem can be
reduced to a purely three-dimensional calculation.

The Regge equivalent of the scalar curvature for a three dimensional
hypersurface has been given by Wheeler \cite{wheeler64}.  The Regge
equivalent of $R=0$ is
\begin{equation}  \label{eqn:regge_ive}
\sum_b L_{ab} \, \varepsilon _{ab} = 0
\end{equation}
where $L_{ab}$ is the edge joining vertex $a$ to vertex $b$, and
$\varepsilon_{ab}$ is the deficit angle (curvature) about
$L_{ab}$. The summation is over all edges $L_{ab}$ which meet at
vertex $a$.

In order to solve this restricted initial value problem, we mirror the
continuum formulation and perform a conformal decomposition.  This
reduces the problem to the solution of the single Regge-Hamiltonian
constraint, equation \eqn{regge_ive}, for the simplicial conformal
factor.  The conformal factor is defined on the vertices of the
lattice, and we perform the decomposition on the lattice edge lengths,
since they correspond closely to the continuum metric.  For the edge
$L_{ab}$ connecting vertices $a$ and $b$, the decomposition takes the
form
\begin{displaymath}
  L_{ab} = \psi^2_{ab} \, \bar L_{ab}
\end{displaymath}
where $\bar L_{ab}$ is the freely chosen base edge length.
Since the conformal factor is defined on the vertices,
we use the second order accurate expression
\begin{equation}
  \psi _{ab} = \frac{1}{2} \left( \psi_a + \psi_b \right)
\end{equation}
to define the conformal factor $\psi_{ab}$ acting on the edge
$L_{ab}$, where $\psi_a$ is the conformal factor at vertex $a$.

So far we have considered only the restricted case of initial data at
a moment of time symmetry --- a purely three dimensional problem.
However, it is possible to use the solution to equation
\eqn{regge_ive} to obtain full axisymmetric two-surface simplicial
initial data.

After solving equation \eqn{regge_ive}, we have a set of physical
edges $\{L_{ab}\}$, which fix the geometry on a simplicial
three-surface.  Using the time-symmetry condition, we can construct
two-slice initial data by creating two identical three-dimensional
lattices using these edges, and then filling in the intervening
four-volume with four-simplices.  This involves introducing a timelike
edge to join equivalent vertices in the two three-geometries, together
with a ``brace'' edge which stretches from one surface to the next,
with one such brace lying above each edge in the lower surface.  This
constructs a set of four-simplices between the two surfaces.  Care
must be taken in the choice of these braces if a particular lattice
structure is desired.  In particular, if we require a lattice amenable
to the Sorkin evolution scheme \cite{committee}.  

Assuming symmetry about the centre of the intervening volume, the
squared lengths $b_{ab}^2$ of the brace edges stretching from one
surface to the next are
\begin{displaymath}
  b_{a^+b}^2 = L_{ab}^2 + \tau_{a^+a}^2 
\end{displaymath}
where vertex $a^+$ lies on the upper surface, and
$\tau^2_{a^+a} = \tau^2 <0$ is the proper time between the two slices,
measured along the timelike edges connecting equivalent vertices.  All
such timelike edges have the same length, and we require that $-\tau^2
<< \min \{L_{ab}^2\}$ in order to obtain a reasonable approximation to
the time-symmetry condition.  This completes the solution of the full
two-surface initial value problem at a moment of time symmetry.

\section{A lattice built from prisms?}
\label{sec:prism}

So far the discussion of Regge calculus has been general, without the
need to restrict attention to one particular lattice.  Before we can
construct a solution numerically, we must choose a particular form of
lattice to work with.  In this section we discuss the application of a
prism-based lattice to the axisymmetric initial value problem in Regge
calculus at a moment of time symmetry.

Much previous work in Regge calculus has been based on simple lattice
structures, often built from prisms. Typical examples include the work
of Lewis \cite{lewis82}, where several $T^3\times R$ cosmological
models were approximated with a hypercubic lattice, and Collins and
Williams \cite{collins73}, where tetrahedra were used to build each
three-geometry, but the world-tube of each tetrahedra was not fully
subdivided into four-simplices.

The principle advantages of this prism-based approach are the ability to
closely model the symmetries of the spacetime of interest, and the
computational simplicity when compared with a simplicial lattice.
Since there are less edges in the lattice, there are less hinges about
which curvature is concentrated, and hence less calculations are
required to obtain the deficit angles within the lattice.   

It is largely for these reasons that the prism approach has dominated
much previous work in Regge calculus.  The major drawback of a
prism-based lattice is that even when all edges have been specified,
the lattice is still not rigid; additional constraints must be
provided to constrain the space of possible edge lengths.  In the
presence of high symmetry, it is possible that natural restrictions
are available to specify the remaining degrees of freedom.  This has
been the case for most previous prism-based applications of Regge
calculus.

The simplicial approach, which constructs three-manifolds from
tetrahedral and four-manifolds from four-simplices, is more natural if
one wishes to model a complex spacetime.  In a simplicial lattice,
once all edge lengths have been specified, the geometry is uniquely
determined. Noting that a simplicial lattice may be obtained by
subdivision of a prism-based lattice, some of the advantages of the
prism construction may be carried over to the simplicial lattice.

\begin{figure}
  \begin{center}
    \epsfig{file=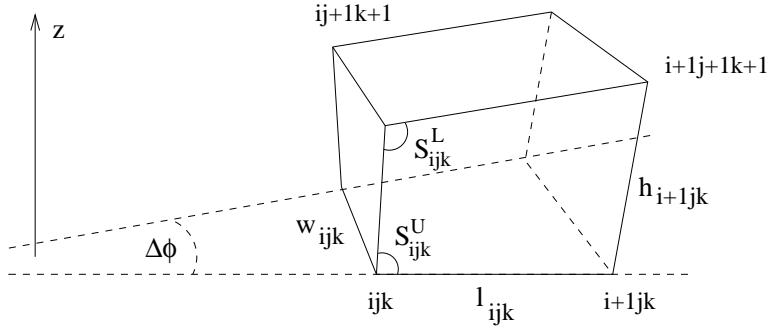,width=4in}
  \end{center}
  \caption{A section of the axisymmetric lattice used by Dubal
    \cite{dubal89a}.  The relation to the global polar co-ordinate
    system is shown, together with the angles used to specify the
    extra degree of freedom.}
  \label{fig:prism}
\end{figure}

For the particular case of axisymmetric non-rotating Brill waves, the
symmetry suggests that a prism-based lattice may be convenient.
Indeed, Dubal \cite{dubal89a} approximated the pure Brill wave
spacetime at a moment of time symmetry with a lattice constructed from
the co-ordinate blocks of the continuum, a typical element of which is
shown in figure \fig{prism}{}.  The desired axisymmetry is built into
the lattice by demanding that there is no variation along the
$\phi$-axis.  The limit as $\Delta \phi$ tends to zero is then taken
to obtain a purely axisymmetric lattice.

Dubal fixed the remaining degrees of freedom in the prisms by
introducing several angles on each $\rho-z$ face, as shown in figure
\fig{prism}{}, and specifying them in terms of the surrounding edges.
The clever assignment chosen by Dubal allowed him to construct Regge
solutions in reasonable agreement with the continuum solution, at
least for low amplitude Brill waves.  Dubal noted several limitations
to his approach, including the relatively poor Regge mass estimates
for ``large amplitude'' ($a \approx \frac{1}{2}$) Brill waves.
Another indication of problems was the low convergence rate of the
prism-based Regge solution to the continuum.  As Dubal notes
\cite{dubal89a}, these problems arise because of the approximation
made in specifying the angles within each prism.  To obtain a better
solution, an improved relation between the angles and prism edges is
required. We argue that the best solution is to abandon the prisms
altogether, and introduce a tetrahedral lattice.

Once the axisymmetric limit ($\Delta \phi \rightarrow 0$) has been
taken, the major difference between a prism-based and a tetrahedral
lattice is the absence of a diagonal edge on the $\rho z$ faces. Each
such face in a prism-based lattice is required, by construction, to be
flat.  However, the pure Brill wave metric \eqn{brill} allows
fluctuations across faces in the $\rho-z$-plane, through both the
conformal factor $\psi(\rho,z)$ and the form function $q(\rho,z)$. A
prism-based lattice is unable to capture such variations.

The most natural solution to this problem is to abandon the
prism-based lattice for one constructed from simplices.  In the
remainder of this paper, we shall follow just such a path, and show
that a tetrahedral lattice resolves all of the problems encountered by
Dubal.

% -----------------------------------------------------------------

\section{A tetrahedral three-geometry}
\label{sec:simplicial}

In this section we construct an axisymmetric, shear-free simplicial
three-geometry, and use it to build time symmetric initial data using
Regge calculus.

\begin{figure}
  \begin{center}
    \epsfig{file=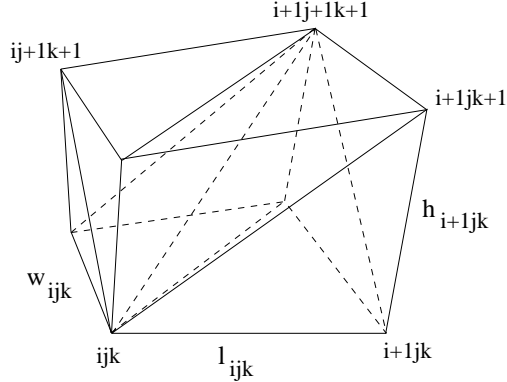,height=2in}
  \end{center}
  \caption{The rectangular prism shown in figure \fig{prism}{}
    subdivided into six tetrahedra.  This involves adding a diagonal
    brace to each face of the prism, together with a body diagonal.}
  \label{fig:tets}
\end{figure}

The simplicial lattice may be obtained by subdividing the basic prism
shown in figure \fig{prism}{}. Each block is divided into six
tetrahedra, introducing three face diagonals and one body diagonal per
vertex.  The result is shown in figure \fig{tets}{}.  The body
diagonal within the prism is denoted $b_{ijk}$, whilst face diagonals
spanning the $\rho$ and $\phi$-axes are written $d^{\rho\phi}_{ijk}$,
and so forth.

To obtain an axisymmetric approximation in the style of the preceding
section, we take the limit as the prism is collapsed along the
$\phi$-axis.  This is a more complicated procedure than in the
previous case \cite{dubal89a}, since we must now deal with the
limiting form of the extra edges introduced in the simplectic
approach, together with their associated deficit angles.

To aid in constructing the lattice, we consider the ``prototype''
axisymmetric metric
\begin{equation}
  \label{eqn:prototype}
  ds^2 = d\rho^2 + dz^2 + \rho^2 d\phi^2.
\end{equation}
This non-physical metric is useful to frame our discussion of the
axisymmetric lattice, and later we will map from this ``prototype'' to
metrics of physical interest.  Figure \fig{tets}{} shows the
subdivision of a co-ordinate block, such that the $l$, $h$ and $w$
edges are locally aligned with the $\rho,z$ and $\phi$-axes
respectively.  We enforce axisymmetry by explicitly setting
\begin{equation}
  w_{ijk} = r_{ijk} \Delta \phi,
\end{equation}
and demanding that there be no variation in the edges along the
$\phi$-axis.  The redundant $j$ index is neglected in the remainder of
this paper.

Consider the behaviour of the lattice in the limit as $\Delta \phi$
approaches zero.  For a single triangle with edge lengths $a,b$ and
$r\Delta \phi$, in the limit as $\Delta \phi\rightarrow0$ we require
that $a=b$.  In general, the manner in which the edges $a$ and $b$
approach this limit is not uniquely defined.  However, as we shall
see, the assumption of axisymmetry uniquely determines the leading
order terms in the expansions about $\Delta \phi=0$.

Assuming that the lattice is both axisymmetric and shear-free, the
$\rho-\phi$ and $z-\phi$ faces of the block must be flat as $\Delta
\phi$ approaches zero, since any deviation from flatness introduces an
asymmetry across the prism.  This allows us to write an expansion in
$\Delta \phi$ for the diagonal edges on these two faces. For the
diagonal edge within the $\rho-\phi$ face we have
\begin{equation}  \label{eqn:diagonal_xy}
  \left\{ d^{\rho\phi}_{ik} \right\}^2 = l_{ik}^2 
  + r_{ik} r_{i+1k}\Delta \phi^2,
\end{equation}
and similarly, the diagonals which span $z\phi$ faces of the prism may
be expanded as
\begin{equation}  \label{eqn:diagonal_yz}
 \left\{ d^{z\phi}_{ik} \right\}^2 = h_{ik}^2 
 + r_{ik}r_{ik+1}\Delta \phi^2,
\end{equation}
in the limit as $\Delta \phi$ tends to zero.

The only remaining edge which spans the $y$-axis is the body diagonal,
$b_{ik}$.  The shear-free and axisymmetric nature of the lattice also
implies that the plane formed by the four vertices $(i,j,k)$,
$(i,j+1,k)$, $(i+1,j,k+1)$ and $(i+1,j+1,k+1)$ is flat as
$\Delta \phi$ tends to zero, and so the body diagonal may be expanded
as
\begin{equation}  \label{eqn:diagonal_xyz}
  b_{ik}^2 = \left(d^{\rho z}_{ik}\right)^2 
  + r_{ik} r_{i+1k+1}\Delta \phi^2,
\end{equation}
to leading order in $\Delta \phi$.

We now turn to the calculation of the deficit angles.  Consider a
tetrahedron consisting of the vertices $(0,1,2,3)$, with corresponding
edges $L_{ab}$, where $a$ and $b$ are vertices of the tetrahedron.
The dihedral angle $\theta^{0123}_{01}$ about the edge joining
vertices $0$ and $1$ within the tetrahedron $0123$ is
\begin{eqnarray}
  \label{eqn:3d_defect}
  \cos \theta^{0123}_{01} =
  \frac{-1} {16A_{012}A_{013}}
  & & \left\{  L_{01}^4 + 2L_{01}^2 L_{23}^2 - L_{01}^2  L_{02}^2
    -L_{01}^2 L_{03}^2 \right. \\
  & & \left. \quad
    - L_{01}^2 L_{12}^2 
    - \ L_{01}^2 L_{13}^2  
    + L_{02}^2 L_{03}^2 + L_{12}^2 L_{13}^2 \right. \nonumber \\
  & & \left. \quad
    - L_{02}^2L_{13}^2 - L_{03}^2L_{12}^2  \nonumber
  \right\}  
\end{eqnarray}
where $A_{012}$ and $A_{013}$ are the areas of the triangular faces
$012$ and $013$ respectively, given by 
\begin{equation}
  \label{eqn:triangle_area}
  16A^2_{abc} = - L_{ab}^4 - L_{ac}^4  - L_{bc}^4
  + 2L_{ab}^2 L_{ac}^2 + 2L_{ab}^2 L_{bc}^2 + 2L_{bc}^2 L_{ac}^2.
\end{equation} 

Equations \eqn{3d_defect} and \eqn{triangle_area} are used to
calculate all dihedral angles within the prism of figure
\fig{tets}{}. Consider, for example, the deficit angle about the edge
$l_{ik}$.  There are six tetrahedra hinging on this edge, so we must
calculate the dihedral angle about $l_{ik}$ in each of these
tetrahedra, after which a power series expansion is taken about
$\Delta \phi = 0$, using equations \eqn{diagonal_xy} ---
\eqn{diagonal_xyz}.  Retaining only the leading power in $\Delta
\phi$, the defect about $l_{ik}$ is
\begin{eqnarray}
  \label{eqn:defect01}
  \epsilon (l_{ik}) & = & \left[
    \frac
    { l_{ik}^2 \left( r_{ik} + r_{i+1k} - 2 r_{i+1k+1}\right) 
      + \Delta_i r_{ik} \left( d_{ik}^2 - h_{i+1k}^2 \right) }
    { 4 l_{ik} A^B_{ik} }  \right. \\
  & & \left. \quad + \frac
    { l_{ik}^2 \left( r_{ik} + r_{i+1k} - 2 r_{ik-1}\right) 
      - \Delta_i r_{ik} \left( d_{ik-1}^2 - h_{ik-1}^2 \right) }
    { 4 l_{ik} A^T_{ik-1} } 
  \right] \Delta \phi,  \nonumber
\end{eqnarray}
where the two triangular areas constituting a $\rho-z$ face are given
by
\begin{eqnarray}
  \label{eqn:areas}
  A^B_{ik} & = & \frac{1}{4} \sqrt{ 2l_{ik}^2 d_{ik}^2 
    + 2l_{ik}^2 h_{i+1k}^2  +  2 h_{i+1k}^2 d_{ik}^2 
    - d_{ik}^4 - l_{ik}^4 - h_{i+1k}^2 }   \nonumber \\
  A^T_{ik} & = & \frac{1}{4} \sqrt{ 2l_{ik+1}^2 d_{ik}^2 
    + 2l_{ik+1}^2 h_{ik}^2  +  2 h_{ik}^2 d_{ik}^2  
    - d_{ik}^4 - l_{ik+1}^4 - h_{ik}^2 },  \nonumber 
\end{eqnarray}
with $\Delta_{i} r_{ik} =r_{i+1k} - r_{ik}$.  

A similar procedure applied to the six tetrahedra about the edge
$r_{ik}\Delta \phi$ yields the deficit angle
\begin{eqnarray}
  \label{eqn:defect02}
  \epsilon (r_{ik} \Delta \phi) = 2\pi
  & - &  \cos^{-1}\left( \frac{ h_{ik}^2+d_{ik}^2 -
      l_{ik+1}^2}{2d_{ik}h_{ik}} \right)          \nonumber \\
  & - &  \ \cos^{-1}\left( \frac{ h_{ik}^2+l_{i-1k}^2 -
      d_{i-1k}^2}{2l_{i-1k}h_{ik}} \right)        \nonumber \\
  & - & \cos^{-1}\left( \frac{ l_{ik}^2+d_{ik}^2 -
      h_{i+1k}^2}{2d_{ik}l_{ik}} \right)          \nonumber \\
  & - &  \ \cos^{-1}\left( \frac{ h_{ik-1}^2+l_{ik}^2 -
      d_{ik-1}^2}{2l_{ik}h_{ik-1}} \right)        \nonumber \\
  & - & \cos^{-1}\left( \frac{ h_{ik-1}^2+d_{i-1k-1}^2 -
      l_{i-1k-1}^2}{2d_{i-1k-1}h_{ik-1}} \right)  \nonumber \\
  & - &  \cos^{-1}\left( \frac{ l_{i-1k}^2+d_{i-1k-1}^2 -
      h_{i-1k-1}^2}{2l_{i-1k}d_{i-1k-1}} \right),  \nonumber
\end{eqnarray}
to leading order in $\Delta \phi$.  The deficit angles about the
vertical edge $h_{ik}$ and the remaining diagonal edge $d^{\rho
z}_{ik}$ are
\begin{eqnarray}
  \label{eqn:defect03}
  \epsilon (h_{ik}) & = & \left[
    \frac
    { h_{ik}^2 \left( r_{ik} + r_{ik+1} - 2 r_{i+1k+1}\right) 
      + \Delta_k r_{ik} \left( d_{ik}^2 - l_{ik+1}^2 \right) }
    { 4 h_{ik} A^T_{ik} }  \right. \\
  & & \left. + \frac
    { h_{ik}^2 \left( r_{ik} + r_{ik+1} - 2 r_{i-1k}\right) 
      - \Delta_k r_{ik} \left( d_{i-1k}^2 - l_{i-1k}^2 \right) }
    { 4 h_{ik} A^B_{i-1k} } 
  \right] \Delta \phi  \nonumber \\
  \label{eqn:defect05}
  \epsilon (d_{ik}^{\rho z}) & = & \left[
    \frac
    { d_{ik}^2 \left( r_{ik} + r_{i+1k+1} - 2 r_{ik+1}\right) 
      - \Delta_{ik} r_{ik} \left( l_{ik+1}^2 - h_{ik}^2 \right) }
    { 4 d_{ik} A^T_{ik} }  \right. \\
  & & \left.  + \frac
    { d_{ik}^2 \left( r_{ik} + r_{i+1k+1} - 2 r_{i+1k}\right) 
      + \Delta_{ik} r_{ik} \left( l_{ik}^2 - h_{i+1k}^2 \right) }
    { 4 d_{ik} A^B_{ik} } 
  \right] \Delta \phi  \nonumber
\end{eqnarray}
where the difference operator $\Delta_{ik}$ is defined by $\Delta_{ik}
r_{ik} = r_{i+1k+1} - r_{ik}$.  The remaining deficit angles are
simply
\begin{equation}
  \label{eqn:zero_defects}
  \epsilon (d^{\rho\phi}_{ik}) = \epsilon (d^{\phi z}_{ik}) =
  \epsilon (b_{ik}) =  0,
\end{equation}
by virtue of the axisymmetric expansions, equations \eqn{diagonal_xy}
--- \eqn{diagonal_xyz}.  Note that the deficit about the face diagonal
$d^{\rho z}_{ik}$, the only face diagonal edge remaining after the
limit $\Delta \phi \rightarrow 0$ has been taken, is not identically
zero.  The fully simplicial model, unlike the prism method described
above, can have non-zero curvature about faces in the $\rho z$-plane.

Evaluating the Regge initial value equation \eqn{regge_ive} on the
simplicial lattice gives
\begin{eqnarray}
  \label{eqn:simplicial_ivp}
\lefteqn{ 0 = l_{ik} \, \epsilon\left(l_{ik}\right)
  + l_{i-1k} \, \epsilon\left(l_{i-1k}\right)}  \ \ \ \ \ \nonumber \\ 
  & & +\ 2 r_{ik} \Delta \phi\, \epsilon\left(r_{ik}\Delta \phi\right)  
  + h_{ik} \, \epsilon\left(h_{ik}\right)
  + h_{ik-1}\, \epsilon\left(h_{ik-1}\right)   \\
  & & +\ d_{ik}\, \epsilon\left(d_{ik}\right)
  + d_{i-1k-1}\,  \epsilon\left(d_{i-1k-1}\right) \nonumber
\end{eqnarray}
to leading order in $\Delta \phi$ at each vertex $(i,k)$ in the
lattice.  This equation may be evaluated once all edge lengths in the
lattice are given, together with the definitions of the deficit angles
in terms of the lattice edges given above.

% -----------------------------------------------

\subsection{Pure Brill wave initial data}

We now consider the case of pure Brill wave initial data, as discussed
in section \ref{sec:pure_brill}.  In order to solve the Regge initial
value equation on the tetrahedral lattice, we perform a conformal
decomposition, similar to the continuum approach.

The lattice is constructed to mirror the cylindrical polar co-ordinate
system in which the continuum Brill wave metric \eqn{brill} is
written.  We obtain the Regge conformal decomposition by integrating
the spacelike geodesics between vertices of the lattice, and assigning
the geodesic lengths directly to the lattice edges.  It is sufficient
to take only the leading order terms in the expansion.

To construct the Regge conformal decomposition for axisymmetric Brill
waves, we must map the lattice edges from the ``prototype metric'',
equation \eqn{prototype}, to the Brill wave metric, equation
\eqn{brill}.  Performing this mapping yields 
\begin{eqnarray}  \label{eqn:conformal_cylindrical}
  l_{ik} & = & \left[ \psi ^2 \, e^q \right] _{i+\frac{1}{2},k} \,
  \Delta \rho _i   \nonumber \\
  h_{ik} & = & \left[ \psi ^2 \, e^q \right] _{i,k+\frac{1}{2}} \,
  \Delta z_k    \\
  d_{ik} & = & \left[ \psi ^2 \, e^q \right] _{i+\frac{1}{2}k +
    \frac{1}{2}} \left(\Delta \rho _i ^2  + \Delta z_k^2\right)^{1/2} 
  \nonumber \\
  r_{ik} \, \Delta \phi & = & \psi ^2_{ik} \, \rho _i \, \Delta \phi
  .\nonumber
\end{eqnarray}
The notation $H_{i+\frac{1}{2},k}$ indicates that the quantity $H$ is
centred between the $(i,k)$ and $(i+1,k)$ vertices.

The boundary conditions applied to the lattice are the same as those
used to construct the continuum solution --- equations
\eqn{brill_boundary1} and \eqn{brill_boundary2}.  They are applied
using a second order power series expansion into the domain.  On the
$\rho=0$ boundary we take
\begin{displaymath}
  \psi = C_0 + C_1\rho + C_2\rho^2 + O(\rho^3),
\end{displaymath}
where the $C_i$ are constants.  Applying reflection symmetry gives
$C_1 = 0$, and evaluating the expansion at $\rho=0,\Delta \rho,2\Delta
\rho$ yields the condition
\begin{equation}
  \label{eqn:prism_hole_inner_r}
  \psi_{0k} = \frac{1}{3}\left( 4\psi_{1k} - \psi_{2k} \right).
\end{equation}
Similarly, the reflection boundary condition at $z=0$ takes the form
\begin{equation}
  \label{eqn:prism_hole_inner_z}
  \psi_{i0} = \frac{1}{3}\left( 4\psi_{i1} - \psi_{i2} \right),
\end{equation}
and at the inner point $\rho=z=0$ an expansion in both $\rho$ and $z$
into the domain suggests a boundary condition  of the form
\begin{equation}
  \label{eqn:prism_hole_inner_rz}
  \psi_{00} = \psi_{01} + \psi_{10}  - \psi_{11}.
\end{equation}
The outer boundary condition, equation \eqn{brill_boundary2}, is
applied to the lattice using a centred finite difference approximation
to the derivative.  It is first converted to pure $\rho$ or $z$
derivatives, to give
\begin{eqnarray}
  \frac{\partial \psi}{\partial \rho} & = & \frac{\rho}{r}
  \left( \frac{1-\psi}{r} \right)
  \quad \mbox{along} \quad  \rho = \rho_{max},  \\
  \frac{\partial \psi}{\partial z} & = & \frac{z}{r} 
  \left( \frac{1-\psi}{r} \right)
  \quad \mbox{along} \quad  z = z_{max}, 
\end{eqnarray}
and then evaluated one point in from the relevant  outer boundary.

\begin{figure}
  \begin{center}
    \epsfig{file=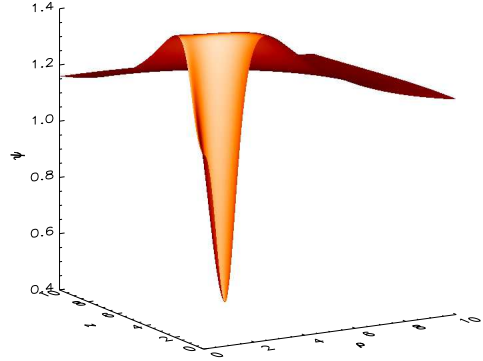, width=0.6\textwidth} 
  \end{center}
  \caption{The conformal factor $\psi$ for a Brill wave of amplitude
    $a=10$  calculated using the simplicial Regge lattice. The
    agreement between this and the continuum solution in figure
    \ref{fig:continuum_brill} is excellent.  Calculations were
    performed using a $601\times 601$ lattice, with the outer
    boundaries placed at $\rho=20$ and $z=20$.}
  \label{fig:simplicial_brill}
\end{figure}

To complete the construction, the lattice spacings $\Delta \rho_i$ and
$\Delta z_k$ are chosen, together with an initial guess for the
conformal factor $\psi$ (taken to be $\psi=1$).  The edge lengths on
the lattice are then calculated using equations
\eqn{conformal_cylindrical}, again using equation \eqn{brill_q} to
specify $q(\rho,z)$.  Finally, the Regge-Hamilton constraint
\eqn{simplicial_ivp} is solved at each vertex, together with the
boundary conditions described above.

The Newton-Raphson method is used to linearize the system, which is
then solved using a sparse storage biconjugate gradient algorithm.
This process is repeated until the desired convergence criteria is
reached; in our case, a fractional change in the conformal factor of
less than 1 part in $10^{-12}$.  Figure \fig{simplicial_brill}{}
displays the solution to equation \eqn{simplicial_ivp} with wave
amplitude $a=10$.  The solution is in excellent qualitative agreement
with the continuum solution shown in figure
\fig{continuum_brill}{}. In section \ref{sec:convergence} we provide a
quantitative analysis of the difference between the two solutions.

We estimate the mass of the initial data shown in figures
\fig{continuum_brill}{} and \fig{simplicial_brill}{} by examining the
fall off of the conformal factor in the asymptotic region.  Assuming
that the conformal factor takes the form 
\begin{equation}
  \psi = 1 + \frac{M}{2r} +  \dots \qquad \mbox{as} \quad r 
  \rightarrow \infty,
\end{equation}
we perform a least squares fit of the this function to each numerical
solution far from the region in which the Brill wave is
concentrated. This asymptotic form for $\psi$ is guaranteed by the
choice of outer boundary condition --- only the value of $M$ is left
undetermined. The mass estimates are shown in table \ref{tab:mass} for
various choices of Brill wave amplitude.  The masses calculated for
the simplicial Regge and continuum solutions agree remarkably well,
and also show excellent agreement with previous calculations.

\begin{table}
\begin{center}
\begin{tabular}{|c|c|c|c|c|}     \hline
  a & $M_S$ & $M_C$ &  $M$ (Alcubierre \etal) \\
  \hline 
  1  & $ 4.7 \times 10^{-2}$  & 
  $4.7 \times 10^{-2}$ & $4.8  \times 10^{-2}$ \\
  2  & $ 1.72\times 10^{-1}$  & $1.72\times 10^{-1}$ & 
  $1.74 \times 10^{-1}$  \\
  5  & $8.77 \times 10^{-1}$ & $8.79\times 10^{-1}$ & 
  $8.83 \times 10^{-1}$ \\
  10 & $3.22$ & $3.22$ & $3.22$ \\     
  12 & $4.84$ & $4.84$ & $4.85$  \\          
  \hline
\end{tabular}
\end{center}
\caption{Mass estimates for the different initial data sets using the
  Eppley form function $q(\rho,z)$. The mass is calculated by
  examining the fall off of $\psi$ in the asymptotic region, for each
  data set --- the simplicial Regge ($M_S$) and continuum ($M_C$)
  solutions.  We also show the results of the previous calculations by
  Alcubierre \etal \cite{alcubierre98}, which are in excellent
  agreement with both the continuum and simplicial Regge solutions.}
\label{tab:mass}
\end{table}

% ------------------------------------------------------------

\subsection{Distorted black hole initial data via Regge calculus}

We now turn our attention to Regge initial data for Brill waves in a
black hole spacetime. Since these waves may be viewed as a distortion
to the Schwarzschild solution, it is natural in to represent them in
spherical polar co-ordinates, as we found in section
\ref{sec:distorted_bh}.

We approximate spherical symmetry in the lattice three-geometry by
aligning the base edges of the lattice with a spherical polar
co-ordinate system.  This implies, for example, that the $z$-axis in
figure \fig{prism}{} becomes the azimuthal $\theta$-axis. For
convenience we also introduce an exponential radial co-ordinate
$\eta$.

In the case of a single black hole with no Brill perturbations, the
face $(h_{ik},r_{ik})$ in figure \fig{tets}{} lies at
constant proper distance from the throat (minimal two-surface), and
the edges $l_{ik}$ stretch from one such face to the next.  For the
remainder of this section the labels $i$ and $k$ are taken to refer to
the radial and azimuthal co-ordinates, respectively. This
identification of the lattice edges greatly simplifies the application
of spherically symmetric boundary conditions.

\begin{figure}
  \fourfigs  
  {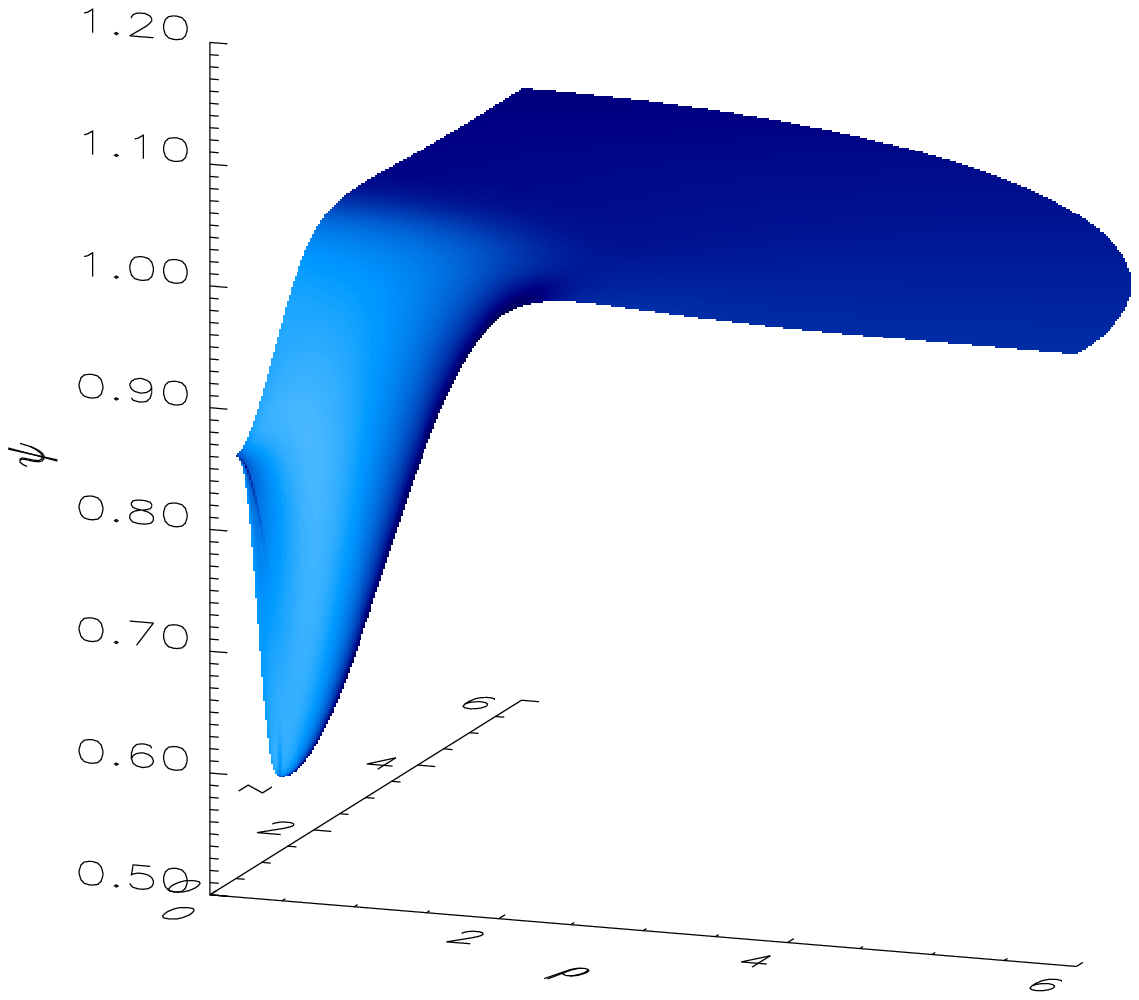}  {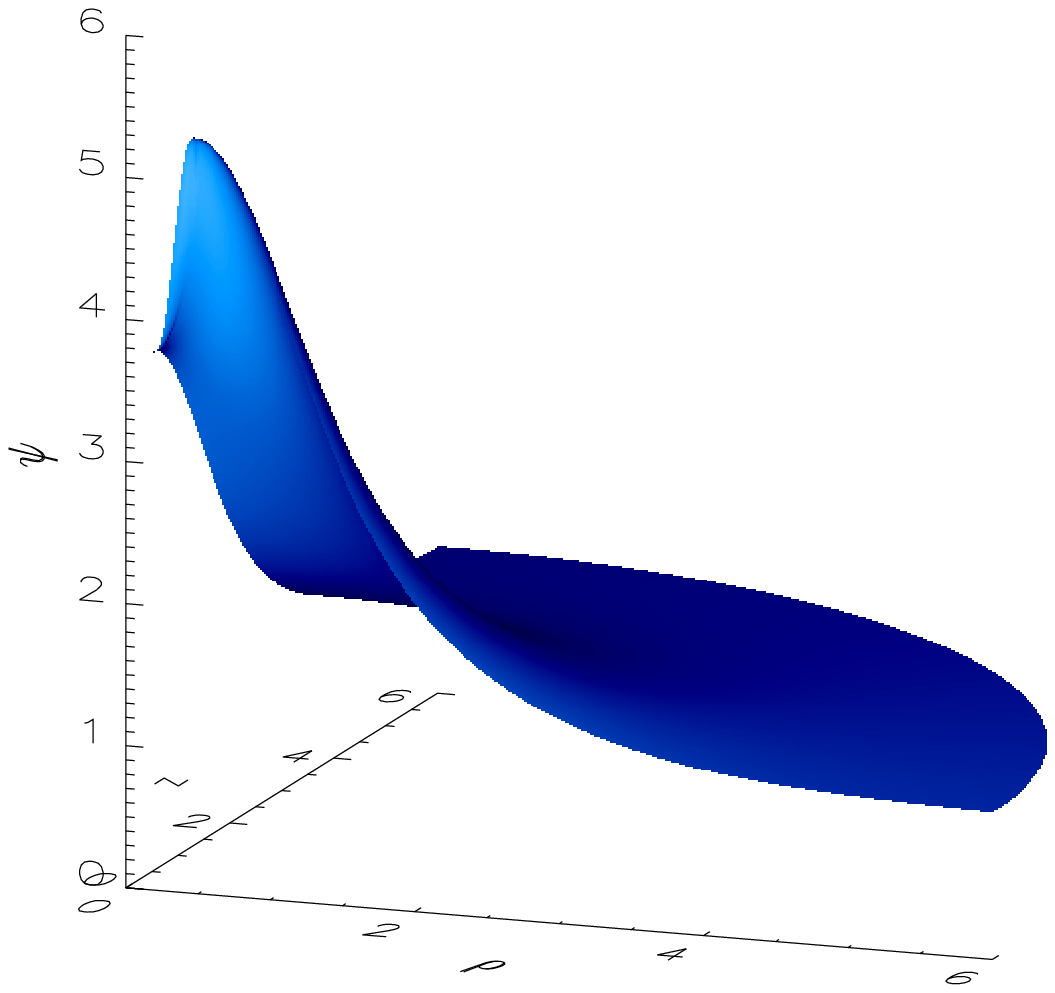}  {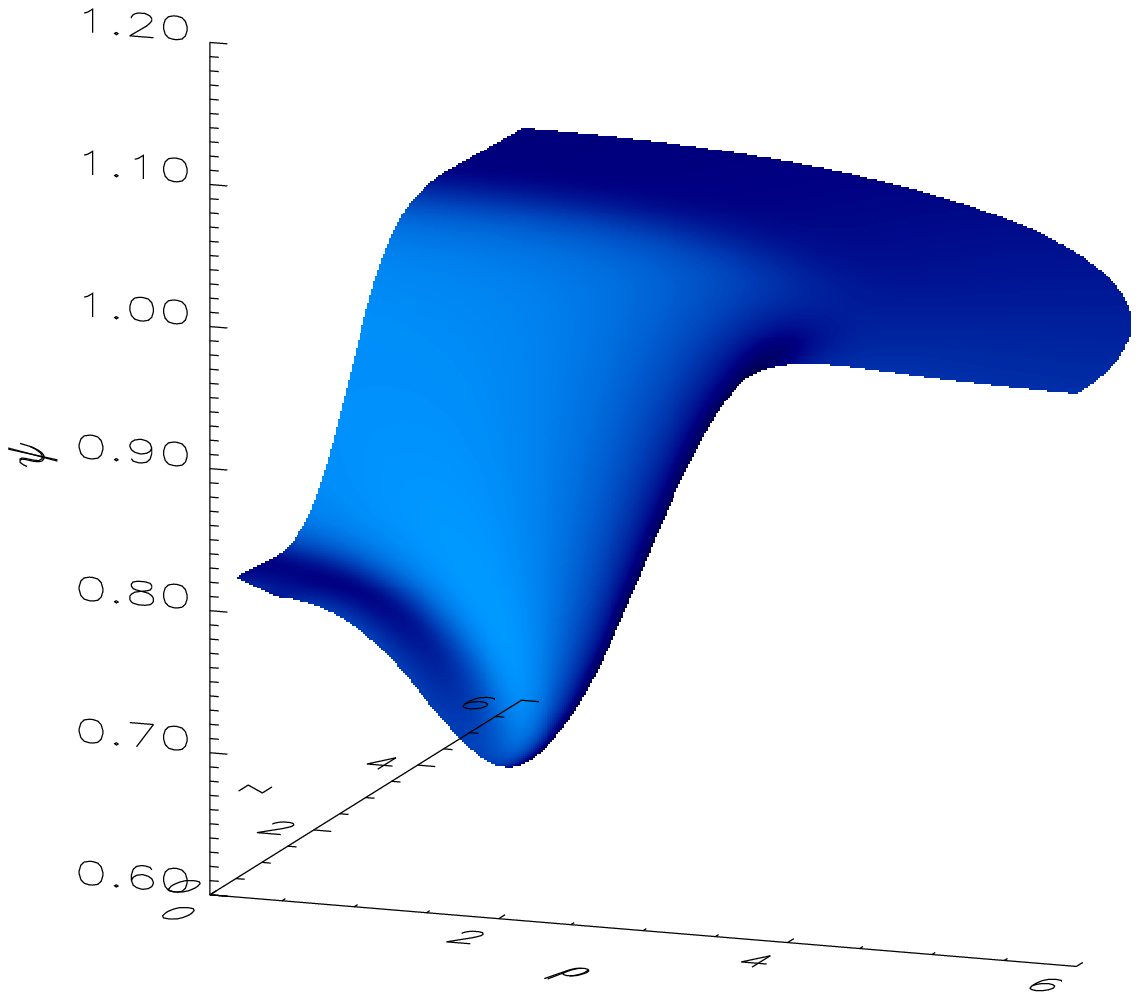}  {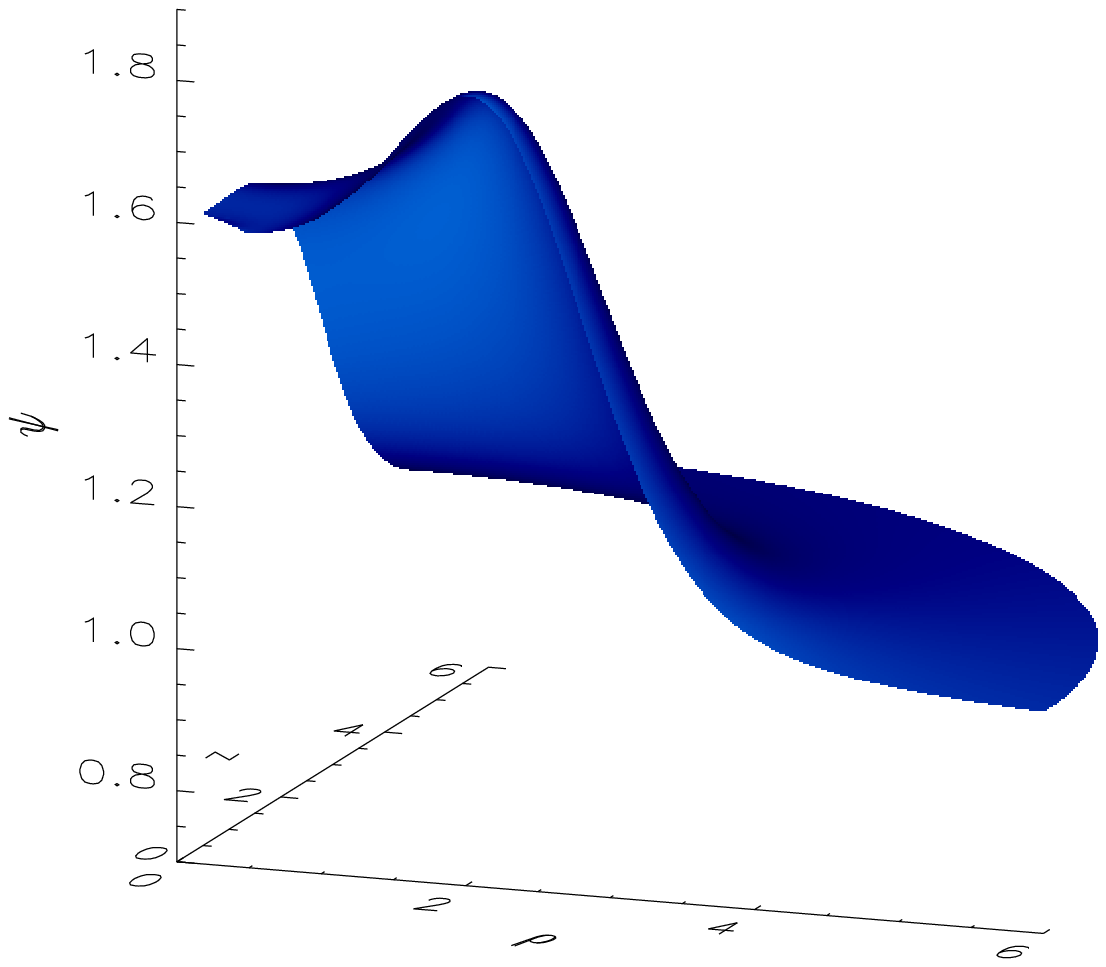}
  \caption{The conformal factor for distorted black hole initial
    data, shown as a ratio of the black hole conformal factor
    ($\psi/\psi_{bh}$). The calculations were performed using the
    simplicial Regge lattice with $601\times601$ vertices, with the
    same choice of parameters as figure \fig{continuum_plots}{}.
    {\sl(a)} $(1,0,1)$, {\sl(b)} $(-1,0,1)$, {\sl(c)} $(1,2,1)$ and
    {\sl(d)} $(-1,2,1)$.  }
  \label{fig:simplicial_plots}
\end{figure}

The metric for black hole plus Brill wave initial data, equation
\eqn{black_hole_brill}, is used to assign the base edge
lengths. Integrating the geodesics between the vertices on each edge,
and retaining only leading order terms, yields a simplicial conformal
decomposition of the form
\begin{eqnarray}  \label{eqn:conformal_spherical}
  l_{ik} & = & \left[ \psi ^2 \, e^q \right] _{i+\frac{1}{2},k} \,
  \Delta \eta _i   \nonumber \\
  h_{ik} & = & \left[ \psi ^2 \, e^q \right] _{i,k+\frac{1}{2}} \,
  \Delta \theta _k    \\
  d_{ik} & = & \left[ \psi ^2 \, e^q \right] _{i+\frac{1}{2}k+
    \frac{1}{2}} \left(\Delta \eta _i ^2  + \Delta \theta_k^2\right)
  ^{1/2} \nonumber \\
  r_{ik} \, \Delta \phi & = & \psi ^2_{ik} \, \sin \theta _k \, \Delta
  \phi, \nonumber
\end{eqnarray}
where we have mapped the lattice edges from the ``prototype'' metric,
equation \eqn{prototype}, to the distorted black hole metric.

The inner boundary conditions are obtained as before, by combining the
continuum inner boundary condition \eqn{hole_boundary1} with a power
series expansion into the domain. The outer boundary uses a second
order finite difference approximation to the Robin condition, equation
\eqn{hole_boundary2}.  The inner $\eta=0$ boundary condition is
\begin{equation}
  \psi_{0k}  =  \frac{1}{3}\left( 4\psi_{1k} - \psi_{2k} \right),
\end{equation}
and similarly, the reflection conditions on at $\theta=0$ and
$\theta=\pi/2$ are
\begin{eqnarray}
  \psi_{i0} & = & \frac{1}{3}\left( 4\psi_{i1} - \psi_{i2} \right) \\
  \psi_{in_\theta} & = & \frac{1}{3}\left( 4\psi_{in_\theta-1} -
    \psi_{in_\theta-2}\right).
\end{eqnarray}
These expansions are combined to give boundary conditions at the two
``corner points'' along $\eta=0$,
\begin{eqnarray}
  \psi_{00} & = & \psi_{01} + \psi_{10}  - \psi_{11} \\
  \psi_{0n_\theta} & = & \psi_{0n_\theta-1} + \psi_{1n_\theta}  
  - \psi_{1n_\theta-1}.
\end{eqnarray}

These boundary conditions, together with the assignment of edge
lengths given by equations \eqn{conformal_spherical}, allow us to
calculate and solve the initial value equation \eqn{simplicial_ivp} as
before, except that we now take the pure black hole solution
\eqn{conformal_black_hole} as the initial guess.  In the following
calculations we use Bernstein's \cite{bernstein93} choice of
$q(\eta,\theta)$, equation \eqn{hole_q}, together with a choice of the
free parameters $(a,b,\omega)$ and the black hole mass $m$.

The solution is shown in figure \fig{simplicial_plots}{}, with the
same set of parameter choices as figure \fig{continuum_plots}.  We
again find excellent qualitative agreement between the continuum
solution and the simplicial Regge initial data.  In the next section
we will attempt to make the comparison more rigorous.

% -----------------------------------------------------------------

\section{Convergence to the continuum}
\label{sec:convergence}

We have seen in the preceding sections that the simplicial Regge
solutions show reasonable agreement with solutions of the continuum
equations.  We now make this comparison more quantitative.

We have at hand two different numerical solutions for a given physical
system and choice of the various parameters; (i) the finite-difference
solution to the continuum equation \eqn{brill_bh_hamiltonian}, and
(ii) the solution of the Regge initial value equation
\eqn{simplicial_ivp}. Since an exact solution is not available for
comparison, the convergence analysis must be based solely on these
equations and corresponding solutions.

Consider the finite-difference solution $\psi_{ik}$ to the
continuum equation.  It represents an approximation to the exact
solution $\Psi$, 
\begin{equation}
  \psi_{ik} = \Psi_{ik} + E_{ik} \Delta ^p + \dots
\end{equation}
where $\Psi_{ik} = \Psi(\eta_i,\theta_k)$, $\Delta$ is a typical grid
scale, and $E_{ij} =\order(1)$.  We expect that $p=2$, since a second
order accurate approximation to equation \eqn{black_hole_brill} was
used to generate $\psi_{ik}$.  In section \ref{sec:distorted_bh} a
convergence test for the pure black hole solution confirmed that the
continuum code is second order accurate.

Now consider the simplicial Regge solution $\psi^R_{ik}$, which we
also expect to differ from the continuum solution by some amount
dependent on the scale length $\Delta$.  By this reasoning we see that
the Regge solution should also differ from the numerical solution of
the continuum equations by some small amount.  If we assume that
\begin{equation}
  \psi^R_{ik} = \psi_{ik} + K_{ik} \Delta ^q + \dots,
\end{equation}
where $K_{ij} = \order(1)$, we can write
\begin{equation}
  \frac {\psi^R_{ik} - \psi_{ik}}{\psi_{ik}} = \frac{
 K_{ik}}{\Psi_{ik}} \Delta ^q + \dots
\end{equation}
where we have neglected only terms smaller than $\Delta^q$.  Noting
that the co-efficient of $\Delta^q$  scales as order unity, we define
the ``error'' between the numerical continuum and Regge solutions as
\begin{equation}
  \label{eqn:truncation}
  e_N = \frac{1}{N}  \sum_{i=0}^{n_\eta}\sum_{k=0}^{n_\theta}  
  \left|  \frac {\psi^R_{ik} - \psi_{ik}}{\psi_{ik}} \right|
\end{equation}
where $N=(n_\eta+1)\times(n_\theta+1)$.  The error $e_N$ may be
calculated directly from the two numerical solutions, and doing so for
a variety of grid scales $\Delta$ yields an estimate of the
convergence rate $q$. 

\begin{figure}[t]
  \begin{center}
    \epsfig{file=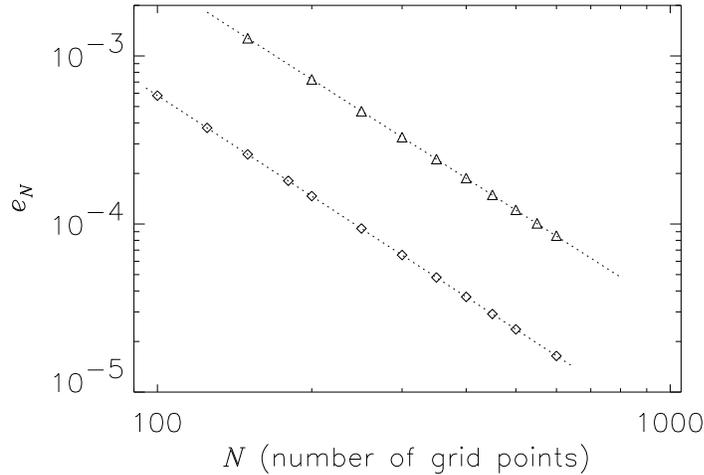,width=4in}
  \end{center}
  \caption{The averaged fractional difference $e_N$ between the simplicial and
    continuum solutions, shown as a function of the number of vertices
    (grid points).  All calculations were performed on an $N\times N$
    grid, for the pure Brill wave space with $a=10$ (triangles), and
    distorted black hole initial data with parameters $(1,2,1)$
    (diamonds).  In both cases, the fractional difference between the
    simplicial Regge and continuum solutions reduces as the second
    power of the grid spacing.  From this we conclude that the
    simplicial Regge solution is a second order accurate approximation
    to the underlying continuum solution. }
  \label{fig:convergence}
\end{figure}

Figure \fig{convergence}{} shows the behaviour of the error $e_N$ as
the number of grid points is increased (hence $\Delta$ decreased) for
both the pure Brill wave and the distorted black hole initial data sets.
The figure suggests a value of $q\approx 2$.  That is, the difference
between the simplicial Regge solution and the numerical solution
$\psi_{ik}$ decreases as approximately the second power of the
discretization scale.  Since the numerical solution $\psi_{ik}$ itself
differs the exact solution $\Psi$ by terms of order $\Delta^2$, we
conclude that the Regge solution is a second order accurate
approximation to $\Psi$, the underlying exact solution of the
continuum equations.

\section{Discussion and further work}

We have successfully used Regge calculus to construct axisymmetric
initial data at a moment of time symmetry. Brill wave and distorted
black hole initial data sets were obtained, and shown to agree well
with the corresponding continuum solutions.

The simplicial lattice --- consisting of tetrahedra in
three-dimensions --- was found to yield solutions in excellent
agreement with the continuum, and displayed none of the problems
encountered with a prism-based Regge lattice.  In particular, we
observed that the simplicial Regge solution converged to the numerical
solution of Einstein's equations as the second power of the lattice
spacing.  From this it was inferred that the simplicial Regge initial
data is a second order accurate approximation to the exact solution.

Even for axisymmetric gravitational waves at a moment of time
symmetry, we have found it necessary to abandon the prism-based
lattice in favour of a simplicial approach.  This indicates that the
future development of Regge calculus as a competitive and robust
alternative to standard numerical techniques will require the use of
fully simplicial lattices.

Work is currently underway to evolve this data set, in
$(2+1)$-dimensions using an axisymmetric adaptation of the Sorkin
evolution scheme \cite{sorkin75,tuckey92,committee}, as well as a
fully $(3+1)$-dimensional evolution.

\ack The author wishes to thank Leo Brewin, Warner Miller and Daniel
Holz for many stimulating discussions on this and related work.
Financial support was provided by the Sir James McNeill Foundation at
Monash University, the Center for Nonlinear Studies at Los Alamos, and
the Theoretical Astrophysics group at Los Alamos through an LDRD
grant.

% --- REFERENCES ----------------------------------------------------
\section*{References}

\gdef\journal#1, #2, #3, #4 {{#1}, {\bf #2}, #3 {(#4)}. }
\gdef\FP{\it{Found.~Phys.}}
\gdef\IJTP{\it{Int. J. Theor. Phys.}}
\gdef\IJMP{\it{Int. J. Mod. Phys.}}
\gdef\GRG{\it{Gen. Rel. Grav.}}
\gdef\PTP{\it{Prog. Theor. Phys.}}
\gdef\AP{\it{Ann. Phys.}}

\end{document}